%% file: main.tex
\documentclass[aps,prl,
twocolumn,superscriptaddress,showpacs,shortbibliography]{revtex4-1}

\usepackage{graphicx}
\usepackage{dcolumn}
\usepackage{bm}
\usepackage{caption}
\usepackage{subcaption}
\usepackage[utf8]{inputenc}
\usepackage[T1]{fontenc}
\usepackage{mathptmx}
\usepackage{hyperref}
\usepackage{cleveref}
\usepackage{float}
\usepackage{color,soul}

\begin{document}

\title{Using a physical model and aggregate data to estimate the spreading of Covid-19 in Israel 
in the presence of waning immunity and competing variants}

\author{Hilla\ De-Leon}
 \email[E-mail:~]{hdeleon@ectstar.eu}
 \affiliation{INFN-TIFPA Trento Institute of Fundamental Physics and Applications, Via Sommarive, 14, 38123 Povo TN, Italy}
 \affiliation{
 European Centre for Theoretical Studies in Nuclear Physics and Related Areas (ECT*),
 Strada delle Tabarelle 286, I-38123 Villazzano (TN), Italy}
 \author{Francesco Pederiva}
\email [Email:~]{francesco.pederiva@unitn.it}
 \affiliation{INFN-TIFPA Trento Institute of Fundamental Physics and Applications, Via Sommarive, 14, 38123 Povo TN, Italy}
 \affiliation{Dipartimento di Fisica, University of Trento, via Sommarive 14, I–38123, Povo, Trento, Italy}

\date{\today}

\begin{abstract}
In more than two years since the COVID-19 virus was first detected in China, hundreds of millions of individuals have been infected, and millions have died.
Aside from the immediate need for medical solutions (such as vaccines and medications) to treat the epidemic, the Corona pandemic has strengthened the demand for mathematical models that can predict the spread of the pandemic in an ever-changing reality.
Here, we present a novel, dynamic particle model based on the basic principles of statistical physics that enables the prediction of the spreading of Covid-19 in the presence of effective vaccines.
This particle model enables us to accurately examine the effects of the vaccine on different subgroups of the vaccinated population and the entire population and to identify the vaccine waning. Furthermore, a particle model can predict the prevalence of two competing variants over time and their associated morbidity.
\end{abstract}

\maketitle


Since SARS-CoV-2 was first identified in China at the end of 2019, hundreds of millions of people have been infected by the virus, and millions have died. In response, several companies and research institutions developed vaccines to help suppress the pandemic and return to pre-corona life. 

By the end of 2020, many countries launched mass vaccination campaigns after the FDA granted emergency authorization approvals (EUA) to two mRNA vaccines - the Pfizer-BioNTech and Moderna vaccines. Even in countries where most people had been immunized, including Israel, Corona cases spiked in the summer of 2021, despite optimism that vaccines would end the pandemic.
There are many possible reasons for the increase in confirmed cases: The removal of most social restrictions in Israel as of March 2020 can be one reason, or the vaccine may be less effective against various variants of the virus (see, for example, Refs. /cite[abu2021effectiveness, fontanet2021sars]), especially with the Delta variant \cite{doi:10.1056/NEJMsr2105280}, or the effectiveness of the vaccine might be deteriorating overtime or any combination of the above.
The unexpected increase in Corona cases following a massive vaccination campaign that began in Israel and was observed later around the world has emphasized the need for dynamic models that can easily be adapted to a changing reality. Hence, in this letter, we show how the combination of both real-time Israeli aggregated data and a physical spatial dynamic model( see Refs.~\cite{de2020particle, de2020statistical} and Methods) can be used to predict the waning immunity of the Pfizer BioNTech vaccine. Additionally, we show (for the Omicron, BA.1, variant) how the model can predict both morbidity and the prevalence of variants in Israel when multiple variants are present (i.e., when the Omicron (BA.1) variant was imported into Israel between December 2021 and January 2022), depending on the degree of population interaction.


We use atomistic-like computer simulations to model the spread of the epidemics. In all simulations, we employed an MC diffusive model as introduced by De-Leon and Pederiva (Refs.~\cite{de2020particle,de2020statistical} and methods) to study the expected effects of vaccinations on the evolution of both the confirmed cases (CC) and severe hospitalizations and to estimate the current efficacy of the vaccine. \footnote{Note that since new severe cases are independent of either testing policy, they consist of a clear and robust indication of the dynamics of disease and the efficacy of the vaccines (with a ten-day shift) in contrast to the number of new CC. Nevertheless, the data analysis requires a division by age accessible only for confirmed and new severe cases, as opposed to Ref.~\cite{De-Leon2021.02.02.21250630}}, In this work, we evaluate the vaccine's efficacy against infection and severe morbidity based on these data and the model. \footnote{In Israel, a severe case of COVID-19 is defined as a SARS-COV-2 confirmed case with >30 breaths per minute, oxygen saturation on room air <9\%, and/or ratio of arterial partial pressure of oxygen to fraction of inspired oxygen <300mm mercury (these also include critical cases - mechanical ventilation, shock, and/or cardiac, hepatic or renal failure), Based on the information accumulated in Israel over the last year, we found that severe and moderate hospitalization takes place about five days following a positive RT-PCR test (see Ref.~\cite{De-Leon2021.02.02.21250630}). Also, based on the current data from Israel, we found that there is a 5-days shift from being hospitalized to becoming a severe case (figure~\ref{fig_severe_SM}), where the number of the new daily severe cases is approximately 0.6 of the new hospitalized five days earlier (all the data was taken from Refs.~\cite{dancarmoz,datagov})}.
The model used here (similarly to what was reported in Refs.~\cite{de2020particle,de2020statistical}) is a "one-way" Ising-model Monte-Carlo such that a susceptible individual (S) can become sick with a daily probability,
	$P_S=\sum_jP_{SI}$, where $P_{SI}$ is a function of the distance between each susceptible individual $(S)$ in the area and all the infected individual $(I)$. Based on the epidemiology data, we assume that an infected individual starts to be infected three days after being infected and then stops being infected after four days (i.e., on the seventh day). In addition, we add a "vaccination" mechanism that switches a certain fraction of particles from the vulnerable state to the immune state. Contrary to other infection models, such as Susceptible Infected Removed (SIR) ~cite{karin2020adaptive, prem2020effect, zhao2020modeling}, this model is a particle model capable of distinguishing between many different age groups and treating them separately under the assumption that infection occurs in the entire population simultaneously.
Furthermore, particle models can be adjusted to accurately examine the differing effects of vaccination on both subgroups and the entire population of vaccinated individuals. This is a straightforward way to estimate the vaccine's effectiveness in the case of multi-variant scenarios (for example, from Nov 2021~\cite{cov}).

	
	In this work, similarly to Refs.~\cite{De-Leon2021.02.02.21250630,de2020particle,de2020statistical} we based on Ref.~\cite{10.7554/eLife.57309} for the Coronavirus epidemiology data and on Refs.~\cite{doi:10.1063/5.0012009,doi:10.1063/5.0011960} for the physical properties of the virus. However, the purpose of the model is to simulate the total morbidity in a particular geographic area, with only the particle density controlling the rate of infection,i.e., the denser the surface is, the rate of pandemic spread will be higher. This parameterization is similar to the low energy constants (LECs) in effective field theory, which encode the unknown high-energy physics and are determined by fitting data to them(e.g., \cite{Hammer:2016xye}), $R_t$, the theoretical reproduction number, encodes the spread of the pandemic in Israel. 
In this work, the control over the surface area, $S$, allows changing the simulation population density, equivalent to a temporary change of the parameter $R_t$, the theoretical reproduction number, with $R_t\sim\frac{1}{S}$, where $R_t$ is determined from the 
 daily change in the number of CC (see Methods for more details). In particular, all simulations are performed, assuming an initial surface area unit of $S=4$ km$^2$ and N=$1.1\cdot 10^4$ particles. Here, we increased/decreased the area as a function of time in all simulations. such that for km$^2$ and $N=1.1\cdot 10^4$ $R_t=1.2$. Periodic boundary conditions decrease broad confinement effects (e.g., the lockdown of an entire province or city) and focus on local infection dynamics within a given area. \footnote{The application of periodic boundary conditions means that we have an infinite number of identical systems; each system is a replica of the others. I.e., suppose a person leaves the simulation surface on one side. In that case, an equal person will enter the surface from the other side.}. 
\
	Based on Ref.~\cite{10.7554/eLife.57309}, this work has identified a list of parameters and corresponding values that describe the population and the infection's kinetics:\footnote{Theses parameters served as input for our simulations when the combination of all the different parameters are eventually expressed at a specific reproduction number. Since, in the end, we are interested in finding a robust correlation between the reproduction number and the particles' density such that we can adjust the simulations to real-time dynamics.
	}. However, the parameter with the greatest effect on the spreading is the virus serial number, i.e., how long it takes for an infected person to become contagious after infection. \cite{alene2021serial,Kim2021.12.25.21268301}:
	\begin{enumerate}
\item For all variants, except for the Omicron: Each sick person is considered contagious between the 3$^{\text{rd}}$ and the 7$^{\text{th}}$ day, i.e the serial number = $5\pm2$ \cite{alene2021serial}.
		\item For the Omicron variant, the serial number =$2.22\pm 1.62$, which is much shorter and affects the rate of pandemic spread \cite{Kim2021.12.25.21268301}.
\end{enumerate}
	
	Similar to Ref.~\cite{de2020particle,de2020statistical}, our basic model is based on the principles of Brownian motion, such that for each day, the population position ($R$) and displacement ($\Delta R)$ are given by:
	\begin{equation}
	R\rightarrow R+\Delta R~,
	\end{equation}
	where $\Delta R=\sqrt{\Delta x^2+\Delta y^2}$ is distributed normally:
	\[
	P[\Delta R] =\frac{1}{2\pi^2\sigma^2_{R}} \exp\left(-\frac{\Delta x^2+\Delta y^2}{2\sigma_{R}^2}\right)~,
	\]
	where $\Delta x$ ($\Delta y$) is displacement in the x-(y-) direction and $\sigma_{R}^2$, the variance, is a function of the diffusion constant, $D$:
	\begin{equation}
	\sigma^2_{R}=2Dt~,
	\end{equation}
	where t=1 day. For a Brownian motion, the diffusion coefficient, $D$, would be related to the temperature, $T$, using the Einstein relation:
	\begin{equation}
	D=\mu k_{\text{B}}T~,
	\end{equation}
	where $\mu$ is defined as the mobility, $k_{\text{B}}$ is Boltzmann's constant
	and $T$ is the absolute temperature. By fixing $T=1$, the diffusion coefficient would be directly related to mobility. In our model, during all the time period, the population is allowed to move with $\sigma_{R}=500$ meter.
	
	The infection process can be described by a Gaussian function of the distance from each contact with another infected person, weighted with a factor that accounts for the sick person's condition and social interaction. \footnote{Since the infection probability is a function of the absolute value of the distance between two people and the standard deviation, several distributions, such as a Gaussian distribution and a Lorentzian distribution, could serve for modeling the infection probability under the assumption that the simulation's dynamics dependents on $\sigma_r$ and not on the distribution's tail. The choice of Gaussian distribution was since this is the typical distribution for thermal systems approaches for equilibrium, e.g., Maxwell Boltzmann distribution.}
	
	\begin{eqnarray}\label{eq_pi}
	P_S&=&\text{int}\left(\sum_{I=1}^{n_{\text{infected}}}P_{IS} \times f(t_I)+\xi\right)=\nonumber \\
	&=&\text{int}\left\{\sum_{I=1}^{n_{\text{infected}}}\exp\left[\frac{\left(R_S-R_I\right)^2}{2\sigma_r^2}\right] +\xi\right\}\nonumber
	\end{eqnarray} 
	
	where:
	\begin{itemize}
		\item $int$ is the integer part
		\item $R_S \left(x_S,y_S\right)$ is the location of the $S^{\text{th}}$ susceptible person and $R_I \left(x_I,y_O\right)$ is the location of the $I^{\text{th}}$ infected person, so $|R_S-R_I|$ is the distance between them.
		\item $n_{\text{infected}}$ is the total number of sick people in the area. 
		\item $\sigma_r$ is the standard deviation (here $\sigma_r=2.4$ meters) 
		{since recent studies show that even a slight breeze can drive droplets arising from a human cough over more than 6 meters
			~\cite{doi:10.1063/5.0011960}}.
		\item $f(t_I)$ is a function that takes into account the infectious capacity of the infected person as a function of time. That is, $f(t_I)$ rises to 1 about five days after infection and decreases after seven days of infection.
		\item $\xi$ is a random number uniformly distributed between 0 and 1, needed for converting to infection probability to a number between 0 and 1.
		\end{itemize}
	
Using our model and aggregated data available from Refs.~\cite{dancarmoz,datagov} in this section, we will estimate the vaccine's efficacy in the summer of 2021 and determine the likelihood of the pandemic spreading in Israel based on the assumption of waning immunity. We have adapted the model to account for Israel's vaccination rate and the presence of effective vaccines. 

Estimating the efficacy of the vaccines against both infection and severe morbidly requires distinguishing between $R_t$ (the theoretical reproduction number of the virus, which corresponds to the density of the population) and $R_e$ (the effective reproduction number of the virus). $R_t$ estimates the number of encounters between carriers and susceptible individuals who would have ended in infection without the vaccines and define the population's dynamics level, where $R_e$ is affected by the vaccination rate and by the vaccine's protection against infection (see Section M.II in methods for more details). 

In this case, we assumed that the theoretical $R_t$ in Israel from July 2021 was increasing over time up to $R_t=2.7$ (see Methods), where this rise is attributed to the removal of some social restrictions prevalent in Israel as well as the dominance of the Delta variant in Israel.

\begin{figure}[h]
\vspace{-0.3 cm}
\begin{subfigure}{.5\textwidth}
 \centering
 \includegraphics[width=1\linewidth]{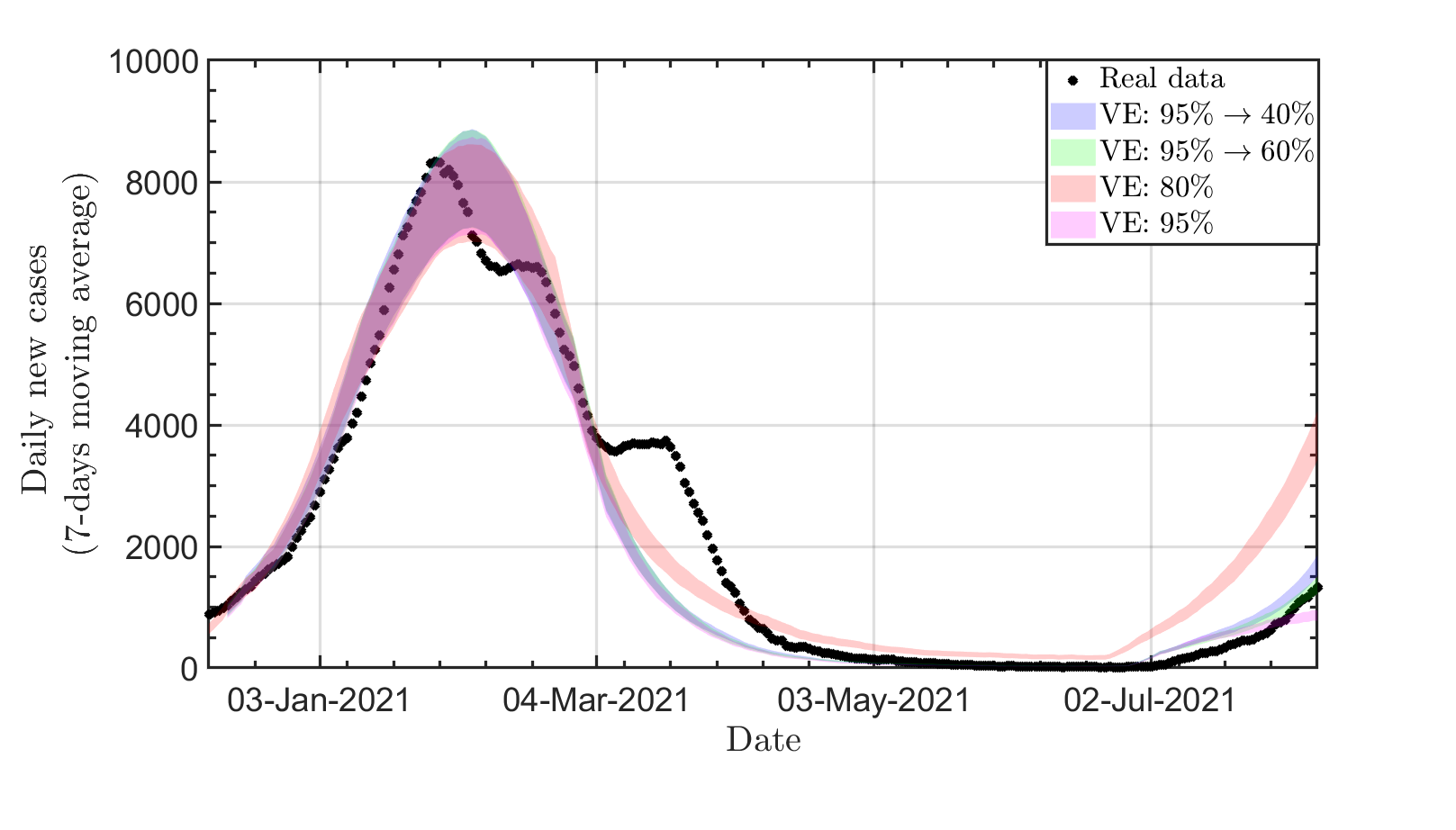} 
 \caption{}
 \end{subfigure}
 \begin{subfigure}{.5\textwidth}
 \centering
 \includegraphics[width=1\linewidth]{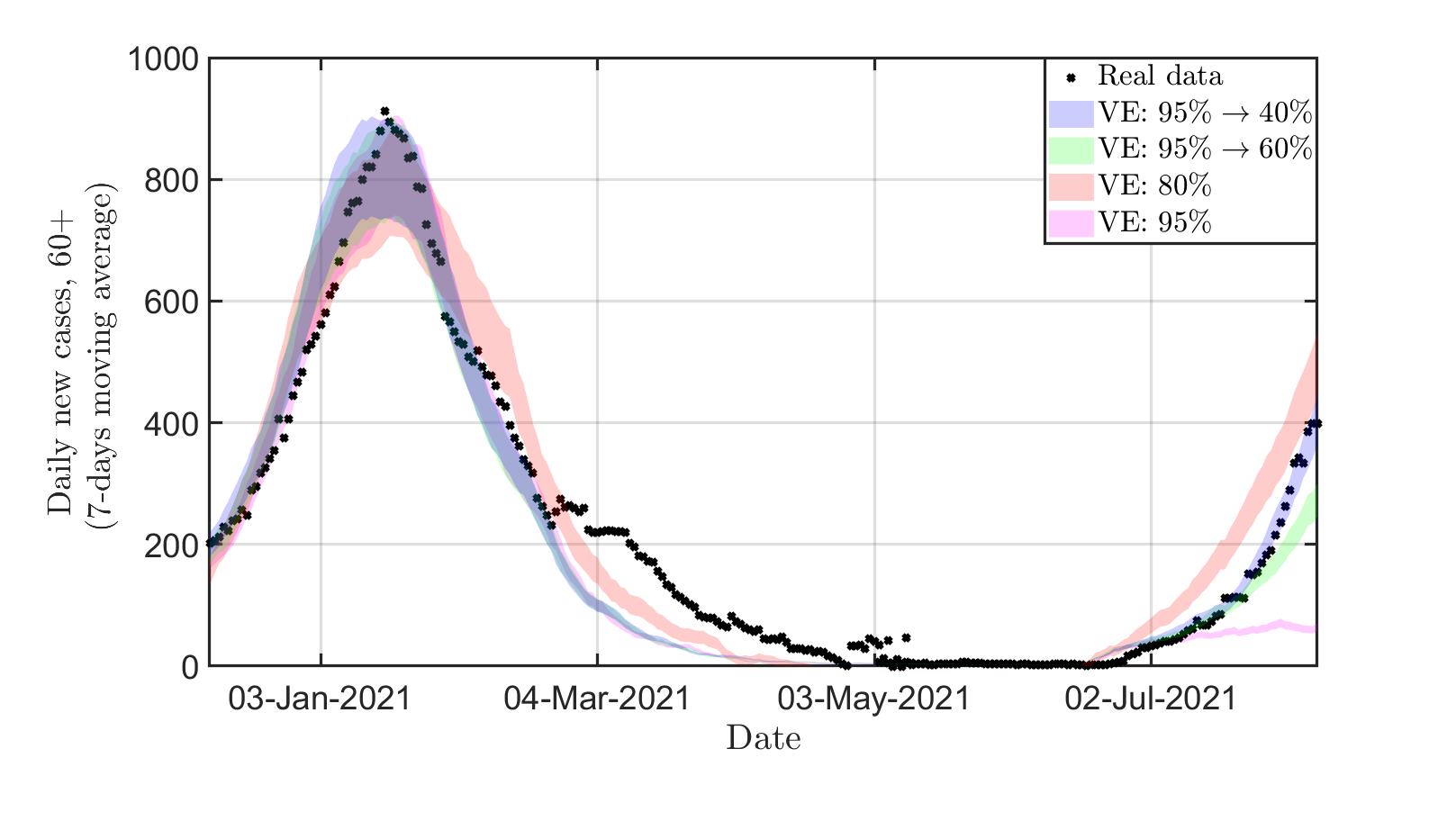} 
 \caption{}
 \end{subfigure}
 \begin{subfigure}{.5\textwidth}
 \centering
 \includegraphics[width=1\linewidth]{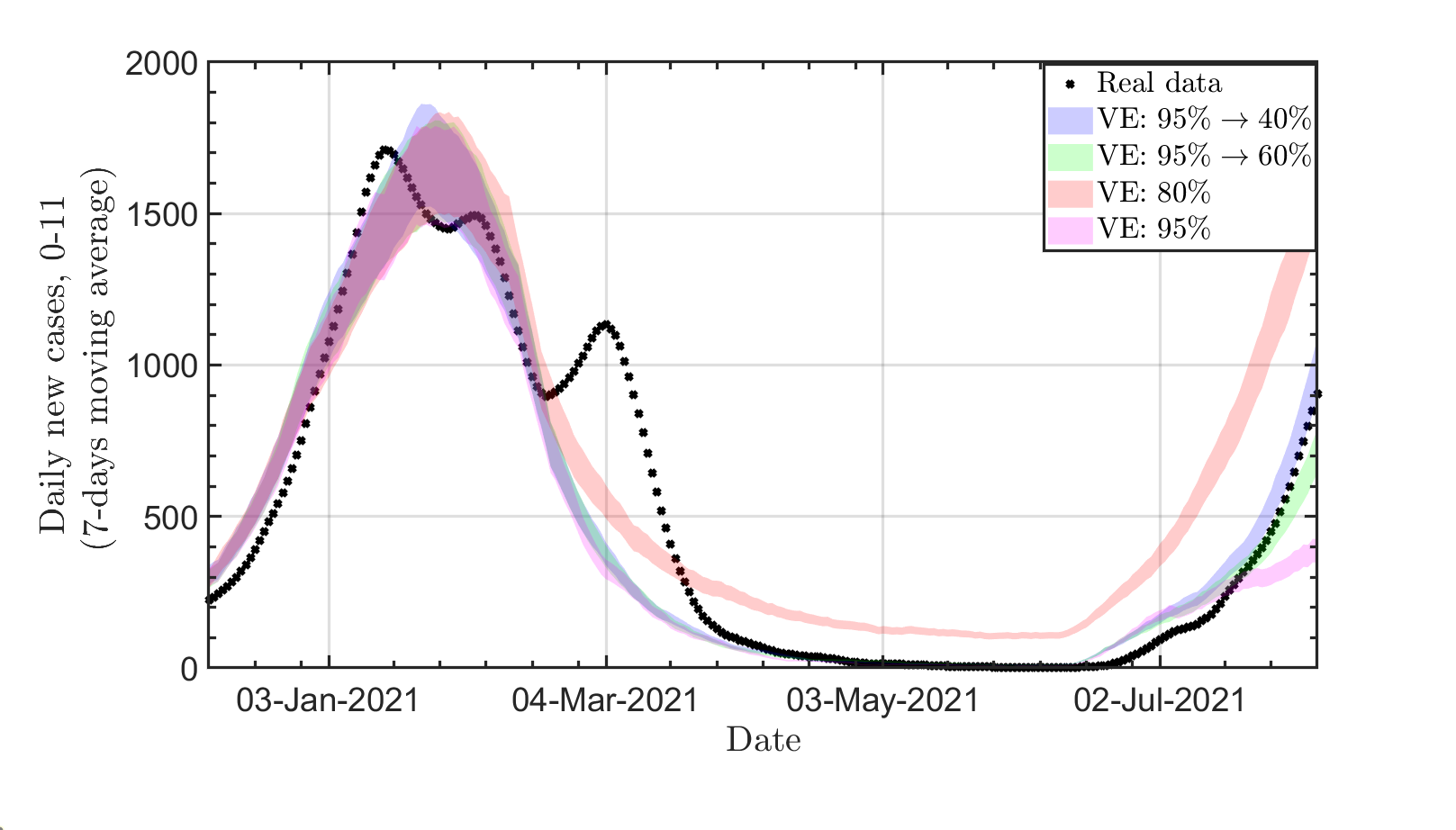} 
 \caption{}
 \end{subfigure}
\vspace{-.3 cm}
 \caption{\footnotesize{\textbf{The daily new CC since January 2021 for three age groups}. The total population (a); people over 60 (b) and people under 11 (c). For all panels: The daily new severe cases form January 2021 of the total population (a) and of people over 60 (b). For both panels: Solid blue band: waning from 95\% efficacy against infection to 40\% after 150 from the second dose; green band: waning from 95\% efficacy against infection to 60\% after 150 from the second dose; red band: 80\% efficacy against infection; magenta band: 95\% efficacy against infection. Crosses: Real data (7-day moving average)
}}
 \label{fig_CC_model}
\end{figure}

\begin{figure}[h]
\begin{subfigure}{.5\textwidth}
 \centering
 \includegraphics[width=1\linewidth]{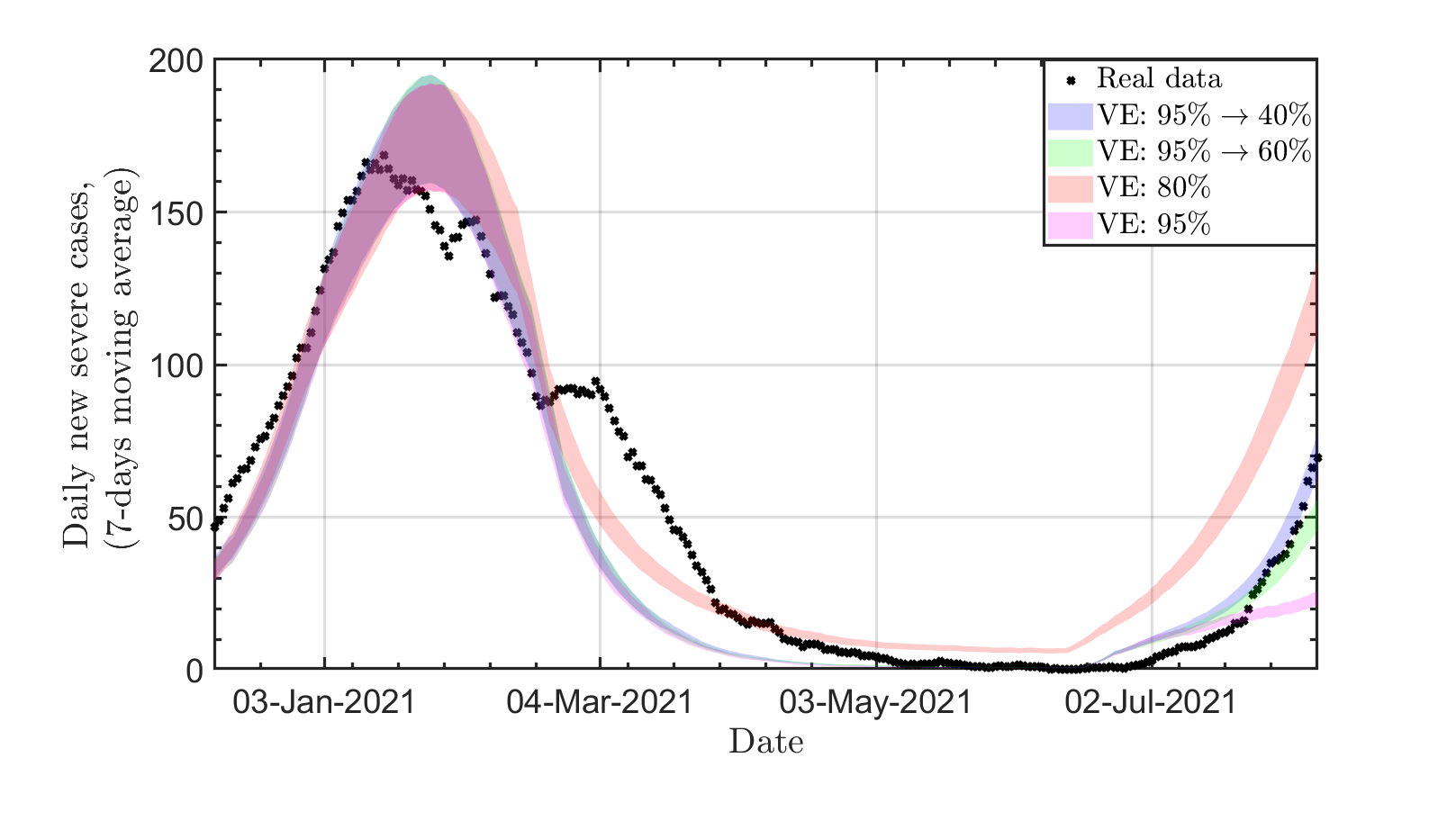} 
 \caption{}
 \end{subfigure}
 \begin{subfigure}{.5\textwidth}
 \centering
 \includegraphics[width=1\linewidth]{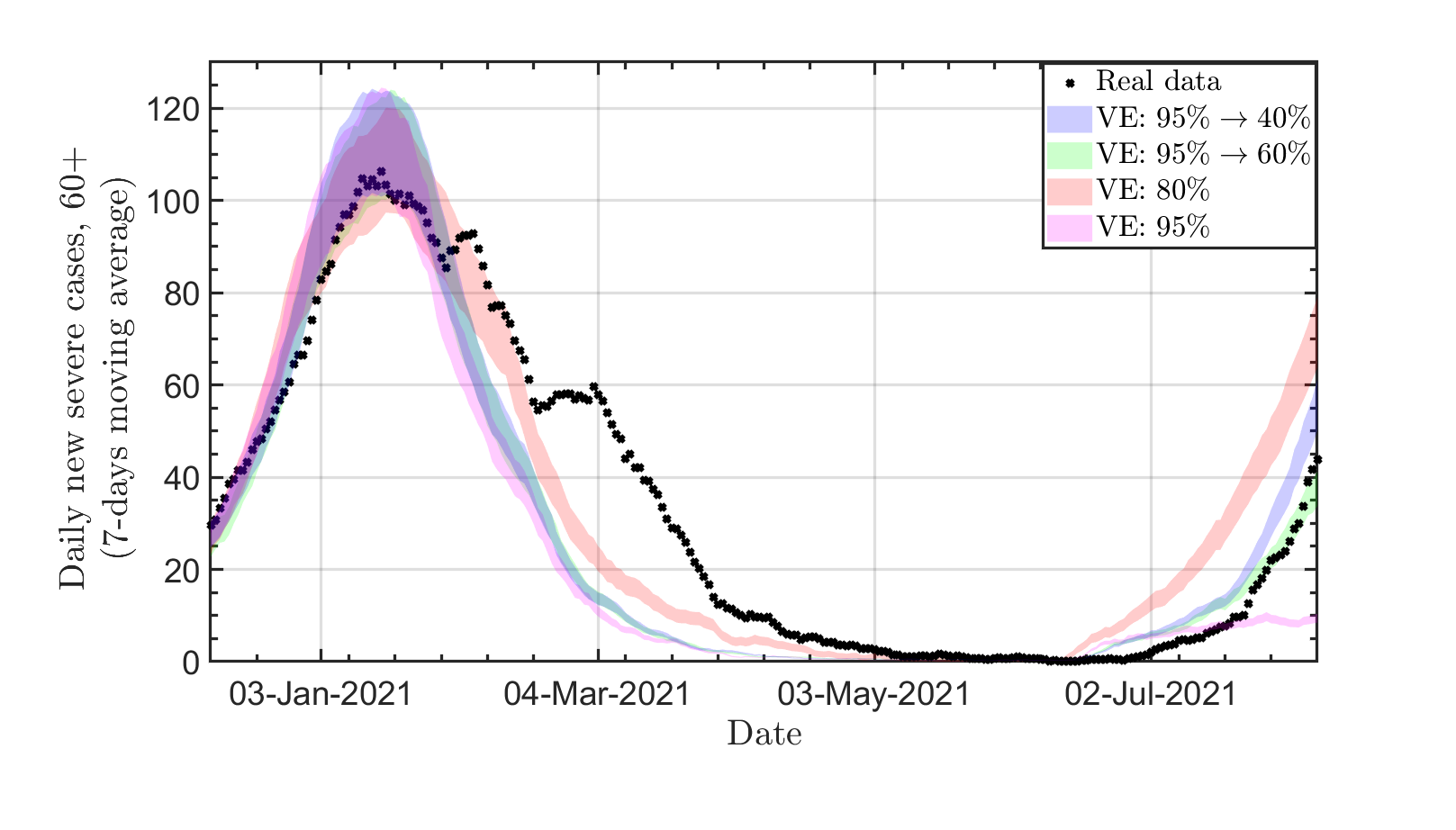}
 \caption{}
 \end{subfigure}
 \caption{\footnotesize{The daily new severe cases form January 2021 of the total population (a) and of people over 60 (b). For both panels: Solid blue band: waning from 95\% efficacy against infection to 40\% after 150 from the second dose; green band: waning from 95\% efficacy against infection to 60\% after 150 from the second dose; red band: 80\% efficacy against infection; magenta band: 95\% efficacy against infection. Crosses: Real data (7-day moving average)}}
 \label{fig_severe_model}
\end{figure}

In this sub-section, we show how using the model and only aggregated data which is accessible to the general public~\cite{datagov}.
Using an MC model, we show in figure~\ref{fig_CC_model} our predictions for the spread of the pandemic in Israel since January 2021 for four different efficacies, all of which assume the same $R_t$ and the current vaccination rate in Israel (see Methods). In figure~\ref{fig_CC_model} we present our predictions for the spread of the pandemic for three age groups: (a) The total population; (b) The elder population (age 60 and up), which can be used as a measure of whether the efficacy of the vaccine is declining over time; and (c) The young population (under 11), which can be used to estimate whether the vaccine can protect even those who are not covered by a vaccine. \footnote{As of July 2021, all of 0-11-year-old in Israel are not protected by the vaccine \cite{datagov,dancarmoz}}. Note that for both figures.~\ref{fig_CC_model} and \ref{fig_severe_model}, we were able to predict the morbidity in Israel from Dec 2020 till the end of July 2021 with a single run. 

Using figure.~\ref{fig_CC_model} one finds that for the case of uniform effectiveness (red and magenta bands in figure.~\ref{fig_CC_model}), there is no correlation in terms of CC between the total number of CC and the number of CC of older people. While the number of CC in the total population is consistent with a model assuming high efficacy against infection (solid magenta band), the number of CC at older ages is not compatible with this efficacy. 

One of the advantages of this model is that it enables us to use time-dependent parameters and, in particular, a time-dependent vaccine's efficacy. Therefore, based on the high percentage of the CC over 60 (the first vaccinators) from the total CC~\cite{datagov,dancarmoz}, for the third and fourth models, we assume high efficacy (95\%) in the first five months after the second dose of the vaccine. Then the efficacy decreases to 60\% (40\%) efficacy against infection, green (blue) band in figure.~\ref{fig_CC_model}). Intestenglty, figure.~\ref{fig_CC_model} shows that the fourth models manage to predict the number of overall and young CC as well as the number of CC aged 60 plus. (Note that based on a comparison between the model and morbidity in Israel in August-December 2020, we assume a 10\% error of the model). 

As can be seen in figure.~\ref
{fig_severe_model}, the model can also predict the total number of daily severe cases and the proportion of severe cases above 60. Similar to Ref.~\cite{De-Leon2021.02.02.21250630}, we assume that the chance of hospitalization is a function of age and workload in the hospital, with suitability for patients in severe condition. Similar to our prediction of the CC, we examined four different scenarios for severe morbidity. For all four models, we assume the same vaccine efficacy against infection as presented in figure.~\ref{fig_CC_model} and that a person's propensity to develop a severe condition after infection is unrelated to their immune status, i.e., protection against severe morbidity is equivalent to protection against infection.

As shown in figure~\ref{fig_CC_model}, both the real data and the model indicate a decline in vaccine efficacy over time. Interestingly, while the confirmed cases for those 60+ are in good agreement with the model assuming a decline to 40/
As a result of the high resolution of this model, it can be used to calculate, on a day-by-day basis, the percentage of CC and severe cases older than 60 who have received their second dose more than 150 days ago. Based on this number and the disparity between the actual data and the model, which assumes waning from 95\% against infection to 40\% (blue band) without additional protection against severe morbidity, one can find that there is a 50\% -70\% protection against severe morbidity for those who are "old" vaccines.

Also, when examining both figure.~\ref{fig_CC_model} and figure~\ref{fig_severe_model}, we found that, based on our estimation of $R_t$ between January and May 2021, it is hard to distinguish between the two (constant) efficiencies in terms of the continuing decline in the number of daily new CC. However, with the increase in daily CC, one finds out that the vaccine's efficacy is a significant factor in daily morbidity and that we can use the same model, only with different vaccine effectiveness, to simulate the morbidity in Israel for more than eight months and to extract the exact waning immunity. 

\footnote{Hence, since the rate of morbidity is a function of both $R_t$ and vaccine efficacy (as shown in the upper panel of figure~\ref{fig_CC_model}), we use the maximum reported efficacy of the vaccine to assess the upper limit of $R_t$ from January 2021 till today (see section M.III in Method), where, in principle, one can assume a lower $R_t$ and lower efficacy of the vaccine for the same morbidity. However, considering a lower $R_t$ will require us to take very low efficacy for early vaccinated adults, which is inconsistent with current studies \cite{Mizrahi2021.07.29.21261317, doi:10.1056/NEJMoa2108891,doi:10.1056/NEJMoa2114228}. }

In June 2021, Israel had experienced its fourth wave of morbidity, while in September 2021, the number of cases declined mainly due to the vaccination of the majority of the Israeli population in the third dose of the vaccine \cite{doi:10.1056/NEJMoa2114255}.
With a large percentage of patients returning from abroad, the number of patients began to rise again at the beginning of December \cite{datagov}.

Therefore, while the morbidity originating in the community does not increase and is caused by the Delta variant, there was a constant flow of new patients afflicted with a new variant whose infectivity is twice that of the delta variant.

Thus, based on our model and the data of the patients who came to Israel between December 1, 2021, and December 15, 2021, we were able to predict:
A. The percentage of Delta cases out of all cases as a function of time.
B. The daily number of cases in Israel for three different $R_t$ scenarios using our assumptions on the prevalence of variants as a function of time (A) and the number of CC in Israel until Jan 11$^{th}$.

Additionally, even though the model uses the immunization rate in Israel as input, it was interesting to compare morbidity trends in Israel and the United Kingdom for December 2021 - January 2022.
\begin{figure}[h]
\vspace{-0.3 cm}
\begin{subfigure}{.5\textwidth}
 \centering
 \includegraphics[width=1\linewidth]{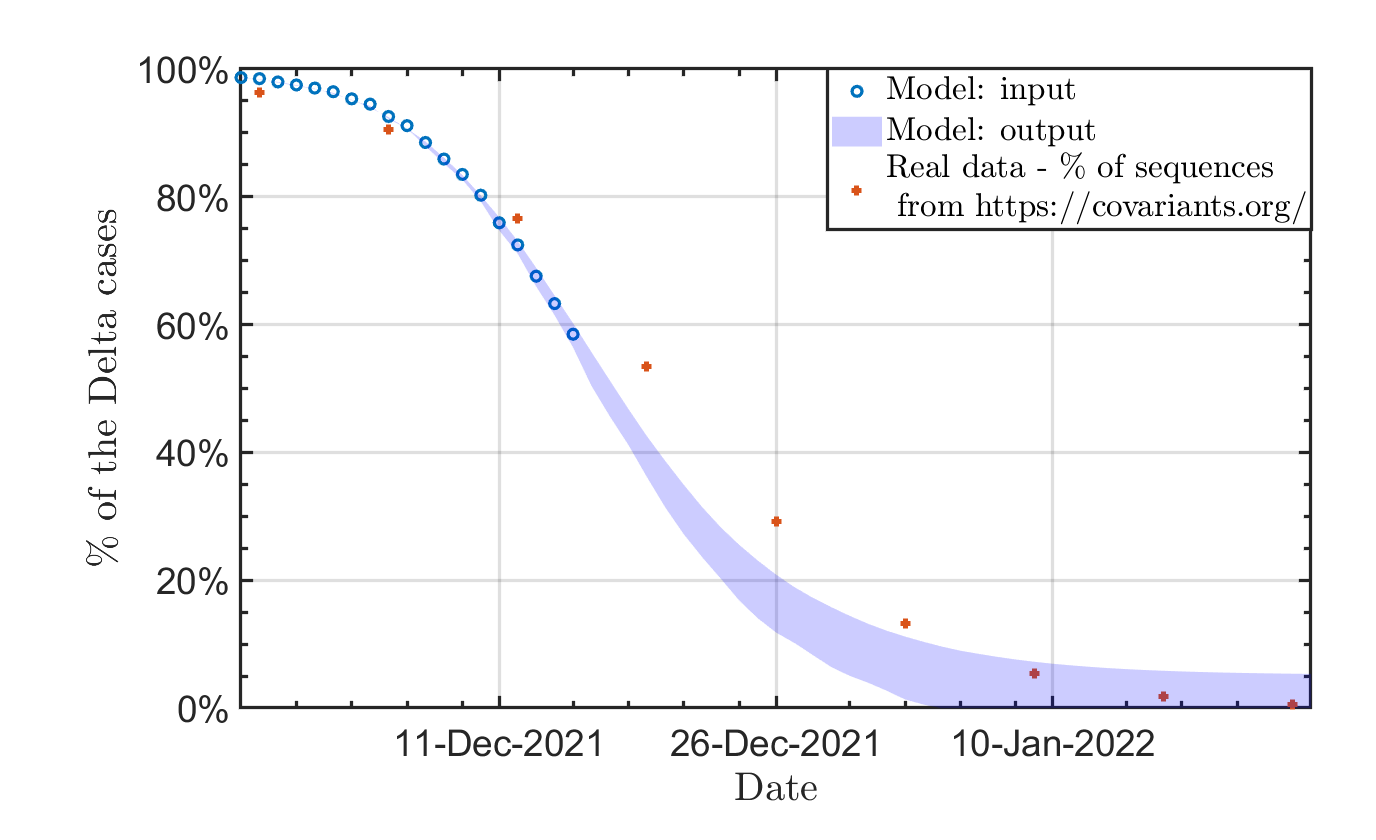} 
 \caption{}
 \end{subfigure}
 \begin{subfigure}{.5\textwidth}
 \centering
 \includegraphics[width=1\linewidth]{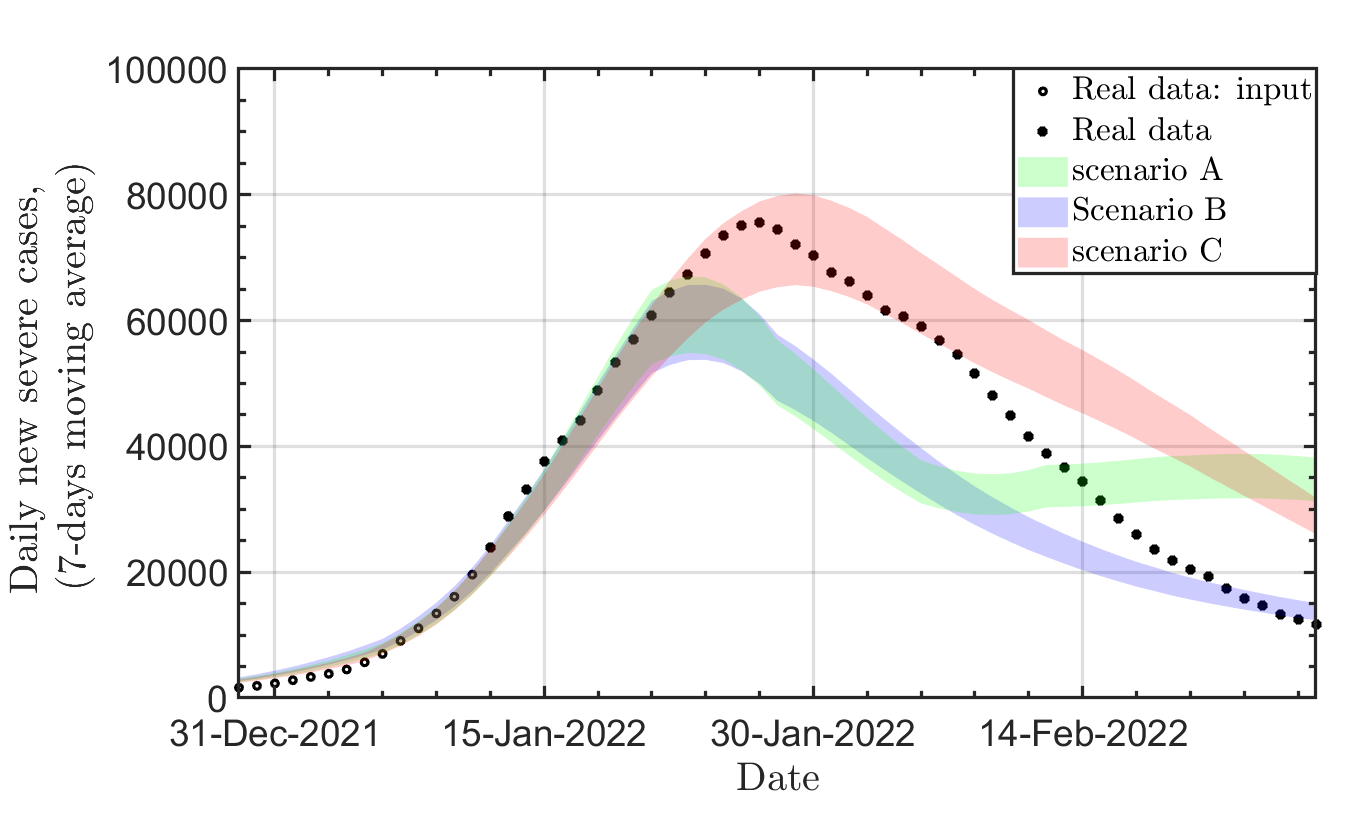} 
 \caption{}
 \end{subfigure}
 \begin{subfigure}{.5\textwidth}
 \centering
 \includegraphics[width=1\linewidth]{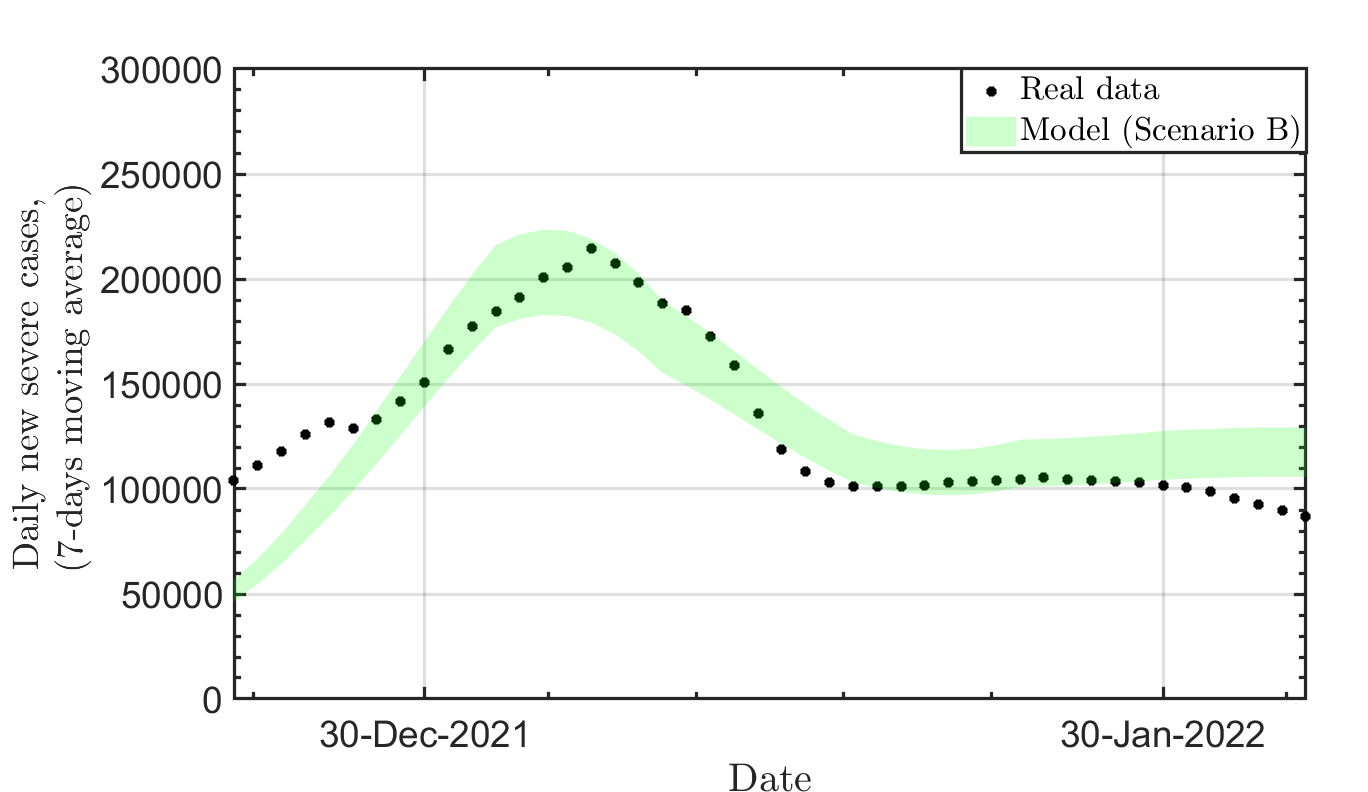} 
 \caption{}
 \end{subfigure} 
\vspace{-.3 cm}
 \caption{\footnotesize{\textbf{Three predictions for the Omicron wave;}(a) The prevalence of variants in Israel . (b) The daily new CC since January 2022 (7-days moving average), based on the The prevalence of variants as shown in (a) and the daily CC until Jun 10$^{th}$, 2022 for three different scenarios of $R_t$ (blue, green and rand bands). the dotes are the real data \cite{datagov}. (c) Modification of scenario B in panel (b) for morbidity in the UK in January 2022: 18-day shift and adjustment to CC's peak. Green band: Model. Dotes: read data from \cite{worldometers}}}
 \label{fig_omicron}
\end{figure}

In figure~\ref{fig_omicron} we show our prediction for the Omicron (BA.1) wave. First, we use the number of daily patients who originate outside Israel.
We assume that between December 1$^{st}$ and December 15$^{th}$, 90\% of those returning from abroad are ill with the omicron virus (BA.1), which is twice as contagious as the Delta variant.
In addition, based on recent studies, we assume that the Delta variant is a faster infection window than the Delta, which will affect the rate of variant BA.1's takeover of morbidity in Israel.
In panel (a) of figure\ref{fig_omicron}, we present our forecast for the distribution of variants in Israel starting in December 2021.
Circles - model's output; Blue band- model's output;
Crosses - actual data, based on the proportion of the total number of sequences (not cases) from \cite{cov}.
In panel (b) of figure\ref{fig_omicron}, we present our prediction for daily CC (moving average of 7 days) for three scenarios which are differ in $R_t$ using the prevalence of variants shown in panel (a) of figure~\ref{fig_omicron} and the daily CC in Israel until Jun 11$^{th}$ . Scenario A (blue line) 2.6 = RT from the beginning of December to January 12$^{th}$, and on January 13$^{th}$ a decrease to Rt = 2.3.
 Scenario B (green line) 2.6 = $R_t$ from the beginning of December to January 12, and on January 12$^{th}$, a decrease to Rt = 2.3. For two weeks, then again, an increase to Rt = 2.6.
Scenario C - (red line) 2.6 = Rt constant throughout the period.
 Dots - Real data \cite{datagov}. In panel (c), We show that even though this model uses the immunization and distribution data of the State of Israel can be adjusted for other countries. Comparing the CC peak of scenario B to the peak of CC in the UK in January and applying a shift of 18 days, we find a good match between the model and the truth data in panel (c). Accordingly, the model fits the case in the UK, where the opening of schools after the Christmas break has resulted in a decline in daily CCs. Thus, in figure~\ref{fig_omicron} it is shown how the model was able to predict the distribution of variants as well as the morbidity rate in Israel since December based only on patient records from outside Israel. Additionally, we found that the model can not only be used to predict Israel's morbidity but also be applied to other countries.
 
We used aggregated data from Israel and a physical dynamic Monte Carlo algorithm to calculate the efficacy of a Pfizer-BioNTech flu vaccine in July (ageist the Delta, B.1.617.2, variant) based on four different vaccine efficacy scenarios and to predict the morbidity in Israel since December 2021 based on two competing variants. The model's main advantage is its extreme flexibility, such that we can test many efficacy scenarios for different age groups with minimal computational cost.
This paper examined how a time- and age-dependent decline in the vaccine's efficacy can affect the pandemic outbreak using a theoretical physical model and aggregate data from Israel. Our model indicates that while the vaccine's effectiveness against infection in most of the population is still high ($~95\%,$, which is consistent with Ref.~\cite{doi:10.1056/NEJMoa2108891}), we detect a decrease in the vaccine's protection against infection for the population vaccinated more than half a year ago. These results are similar to the findings of Ref.~\cite{Mizrahi2021.07.29.21261317}, which found (based on a studied population of more than 1.3 million people) that the risk for infection against the current Delta variant was significantly higher for people receiving an early vaccination compared to those vaccinated later. Also, our finding indicates that during July 2021, there was a decrease in the vaccine's efficacy against severe morbidity for early elderly vaccinators. Still, it is less severe than the decrease in protection against infection. Also, this work shows that the model is very exact even in a complex scenario where there is more than one variant in which each variant has unique features (such as a serial number).
Using our model, we are not only able to predict morbidity but also the distribution of variants over time.
 
This study aims to demonstrate that using accurate models, we can detect vaccine efficacy and variant distribution over time by using \textbf{only} aggregated data. Over the past two years, we have learned that reality is rapidly evolving and that a dynamic model that can be easily adjusted to a changing environment is mostly needed.
Hence, this paper demonstrates that a particle model that uses statistical physics, Brownian motion, and other basic principles can be used to model the propagation of the corona and even make decisions during the third wave of morbidity in Israel (see, for example: \url{ https://corona-analysis.huji.ac.il/2021#january21}).

We thank D. Gazit, Y.Ashkenazy, and R. Calderon-Margalit of the Hebrew University COVID-19 pandemic monitoring team for fruitful discussions. HDL thanks Barak Raveh and Dvir Aran for fruitful discussions.
\pagebreak
		\setcounter{subsection}{0} 
		\setcounter{equation}{0} 
		\setcounter{section}{0} 
		\setcounter{figure}{0}
		\renewcommand{\thesection}{M}
		\renewcommand{\theequation}{M-\arabic{equation}}
		\renewcommand{\thesubsection}{M.\Roman{subsection}}
		\counterwithin{figure}{section}
		\input{methods}

\bibliography{ref}
\bibliographystyle{unsrt}

\end{document}

%% file: methods.tex
\section{Methods}
All simulations simulate the morbidity situation in Israel since the beginning of December 2020 when on December 21, the State of Israel began vaccinating the entire population over 16. The first to be vaccinated were those aged 60+ followed by those aged 50+ and so on (see Fig. S-1). Therefore, the model includes a division by the age of the verified cases as follows:
The simulation consists of 11,000 particles, where each particle has a number from 1 (youngest) to 11,000 (oldest). Since in Israel 17\% of the patients were above the age of 60, we assumed that the highest 17\% numbers in the model represent the population over the age of 60 in Israel, which were the first to be vaccinated in the simulation (and so on to all the age groups) (the daily vaccination rate was taken from Refs.~\cite{datagov,dancarmoz}.

Using the model, we can analyze the spread of the epidemic and its morbidity in Israel by examining the vaccine's efficiencies in preventing infection and morbidity and comparing it to Israel's actual data. In all the simulations, we assumed some protection against infection, which ranges from 40\% to 95
\% after 21 from the date vaccination. Also, we assumed that vaccinated individuals' infectivity is~ 50\% (infected vaccinated patients are less infected than unvaccinated patients). In addition, we examined, using the model, the vaccine's protection against severe morbidity as a function of the time elapsed since the vaccination. The visual comparison between the model and the actual data from Israel for both confirmed cases and severe  morbidity enabled us to evaluate (with ~10\% uncertainty) the vaccine's effectiveness against infection and morbidity.

\subsection{Real time dynamic}
For the model, the important parameters are just the transition probabilities, vaccine effectiveness and rate of population vaccination \cite{dancarmoz,datagov}, as well as the effective reproduction number of the virus,$R_e$. The daily $R_e$ that is used as an input to the simulation is the weekly averaged 4-day growth factor in Israel, until the onset of the third lockdown (January 8, 2021). This is done using the relation between $R_e$ and the particle density (i.e., the area where the simulation occurs, denoted by $S$) such that:
\begin{equation}
 R_e\sim\frac{1}{S}~,
\end{equation}
since $R_e$ represents the number of potential encounters in the population, the area reduction is equivalent to increasing the population density and consequently increasing $R_e$.
\begin{figure}[h!]
 \centering
 \includegraphics[width=1\linewidth]{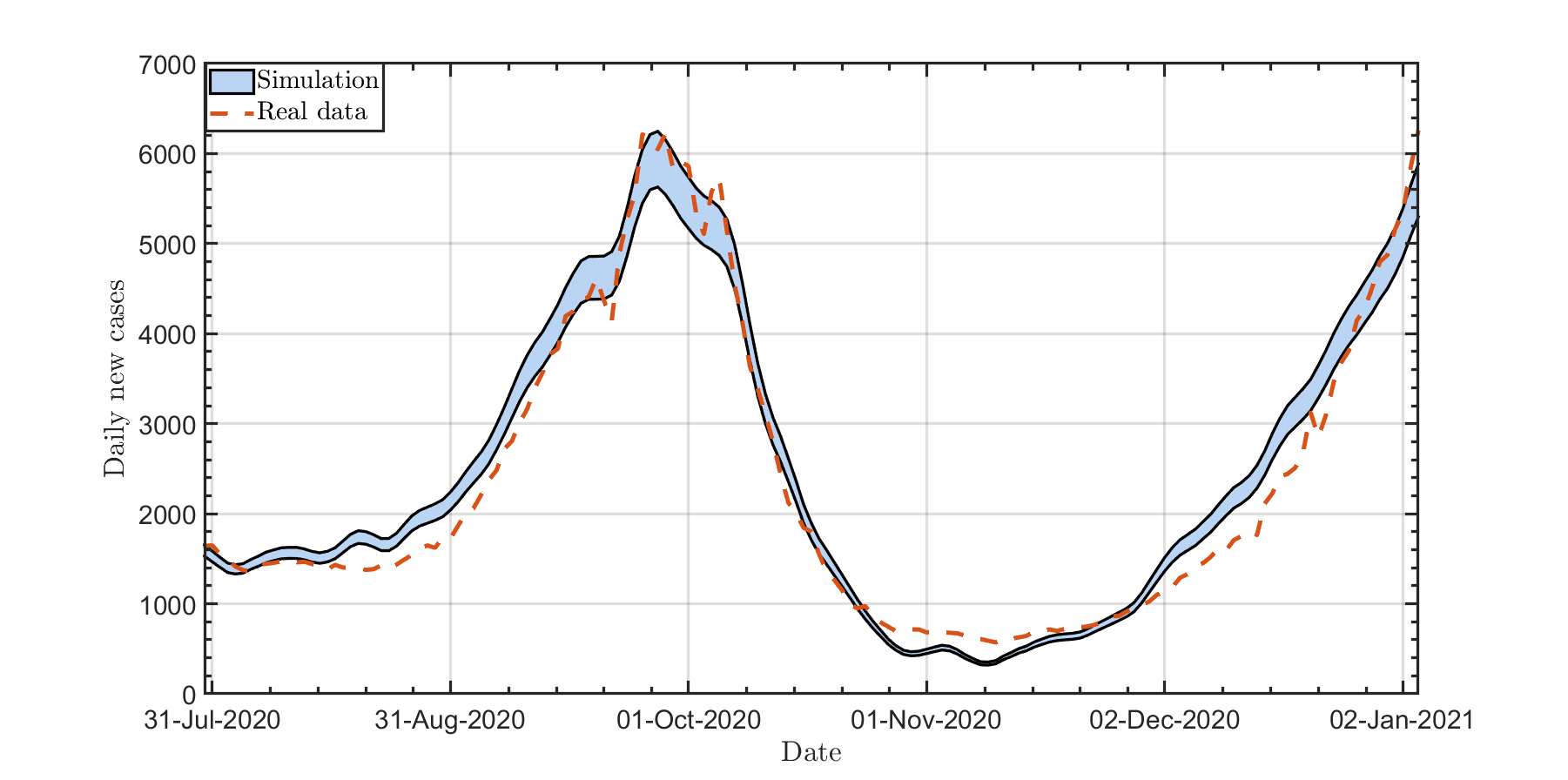}
 \caption{\footnotesize{The daily new confirmed cases between Aug 1st and Jan 6th, 2021. Solid line: simulation. Dashed line: real data~\cite{datagov,dancarmoz} }}
 \label{fig:real}
\end{figure}

Figure \ref{fig:real} shows the agreement of the simulated daily new cases to the reported data in Israel from August 1, 2020, till January 6, 2021, based on the effective reproduction number, $R_e$ in Israel. This bolsters the model's validity and enables us to use it in this work for future predictions.

However, given the vaccination campaign in Israel, a distinction is required between $R_t$, the theoretical effective reproduction number of the virus, and $R_e$ (effective reproduction number of the virus). $R_t$ represents the number of encounters between carriers and healthy people who would have ended in infection without the vaccines and defines the population's dynamics level, where $R_e$ is affected by the vaccination rate and the vaccine's efficacy against infection.

Although it is not trivial to estimate $R_t$, we can assume that for the total population:
\begin{equation}
R_e \approx R_t\times \left(1-N_{vac}\right)= R_t\times \left(1-n_{vac}\times vac_{\text{efficacy}}\right),
\end{equation}
where:
$n_{vac}$ is the percentage of vaccinated in the population and $vac_{\text{efficacy}}$ is the efficacy of the vaccine against infection. 

Therefore, since $n_{vac}$ is known and both the model and the data are age dependant, the correct combination of $R_t$ and $vac_{\text{efficacy}}$ should restore not only the total number of confirmed cases but also the total number of confirmed cases for each age group separately. 

From January 8, 2021, until February 18, 2021, when the Israeli government imposed a national lockdown, the population compliance is simulated such that the population behavior corresponds to $R_t = 1.1$, i.e., each carrier meets on average 1.1 people that would have been infected without vaccines. From February 18 until today, we used In the product:
$R_e\times\frac{1}{1-n_{\text{vac}}}$ as an upper limit for our estimation of $R_t$.

\begin{figure}[h!]
 \centering
 \includegraphics[width=1\linewidth]{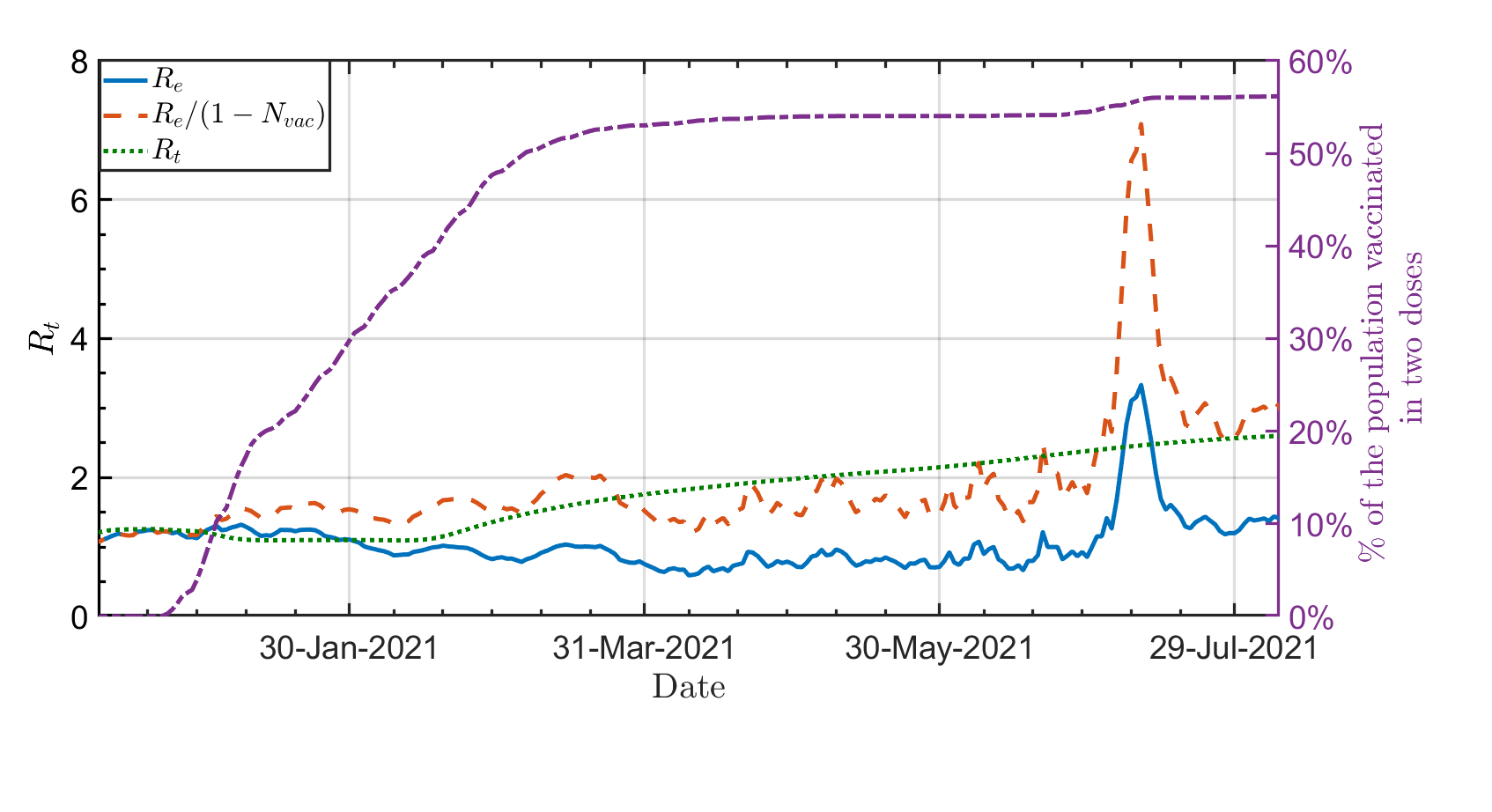}
 \caption{\footnotesize{The confirmed cases effective reproduction number of the virus, $R_e$ , as a function of time. Sold line: $R_e$, based on the daily confirmed cases in Israel; Dashed line:$R_e\times\frac{1}{N_{\text{vac}}}$;Dotted line: $R_t$ =1.1 at lockdown and has been rising gradually since then ; Dotted-dashed line: the percentage of the Israeli population vaccinated in two doses \cite{dancarmoz,datagov}}}
 \label{fig:Rt}
\end{figure}

Figure \ref{fig:Rt} shows the real effective reproduction number, $R_e$, (based on the daily confirmed cases in Israel), which is a result of both the population behavior ($R_t$ ) and the effect of vaccines (solid line). 
\subsection{Moderate and severe daily admissions as an indicator for the state of the pandemic in Israel}
\begin{figure}[h!]
 \centering
 \includegraphics[width=1\linewidth]{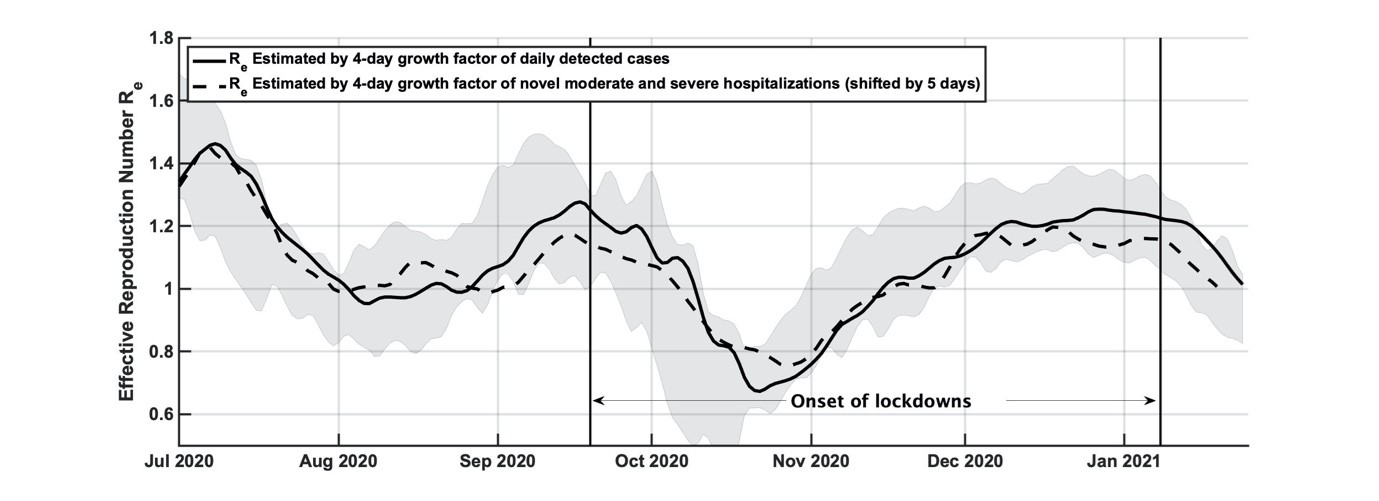}
 \caption{\footnotesize{Solid line: Effective reproduction number estimated by the 4-day growth factor of either detected cases (full line), with its standard deviation (shaded area); Dashed line: moderate and severe new daily hospitalizations, shifted back 5 days to maximally correlate to the detected cases.}}
 \label{fig:SM_cases}
\end{figure}

Based on previous work \cite{De-Leon2021.02.02.21250630}, we use moderate and severe daily admissions as a measure for the disease in Israel. This is justified by the fact that there exists a correlation between this parameter and the population infection rates. This is exemplified in Fig.~\ref{fig:SM_cases}, which showed that there is an agreement between the effective reproduction number estimated by the 4-day growth factor of either detected cases (full line), with its standard deviation (shaded area), or using the moderate and severe new daily hospitalizations (dashed line). Maximal correlation is achieved by shifting back the Re of the newly confirmed cases by five days. As confirmed cases are delayed by five days on average from infection in Israel, this is consistent with a 10-day average deterioration time from infection to moderate or severe condition. As a result, this parameter is a clear and robust indicator for a COVID-19 severe case that has the advantage of being evident quite promptly after infection but does not depend on detection or population test compliance. Thus, it is ideal as an objective measure of disease dynamics and the effect of the vaccine on the outbreak.

\begin{figure}[h!]
 \includegraphics[width=1\linewidth]{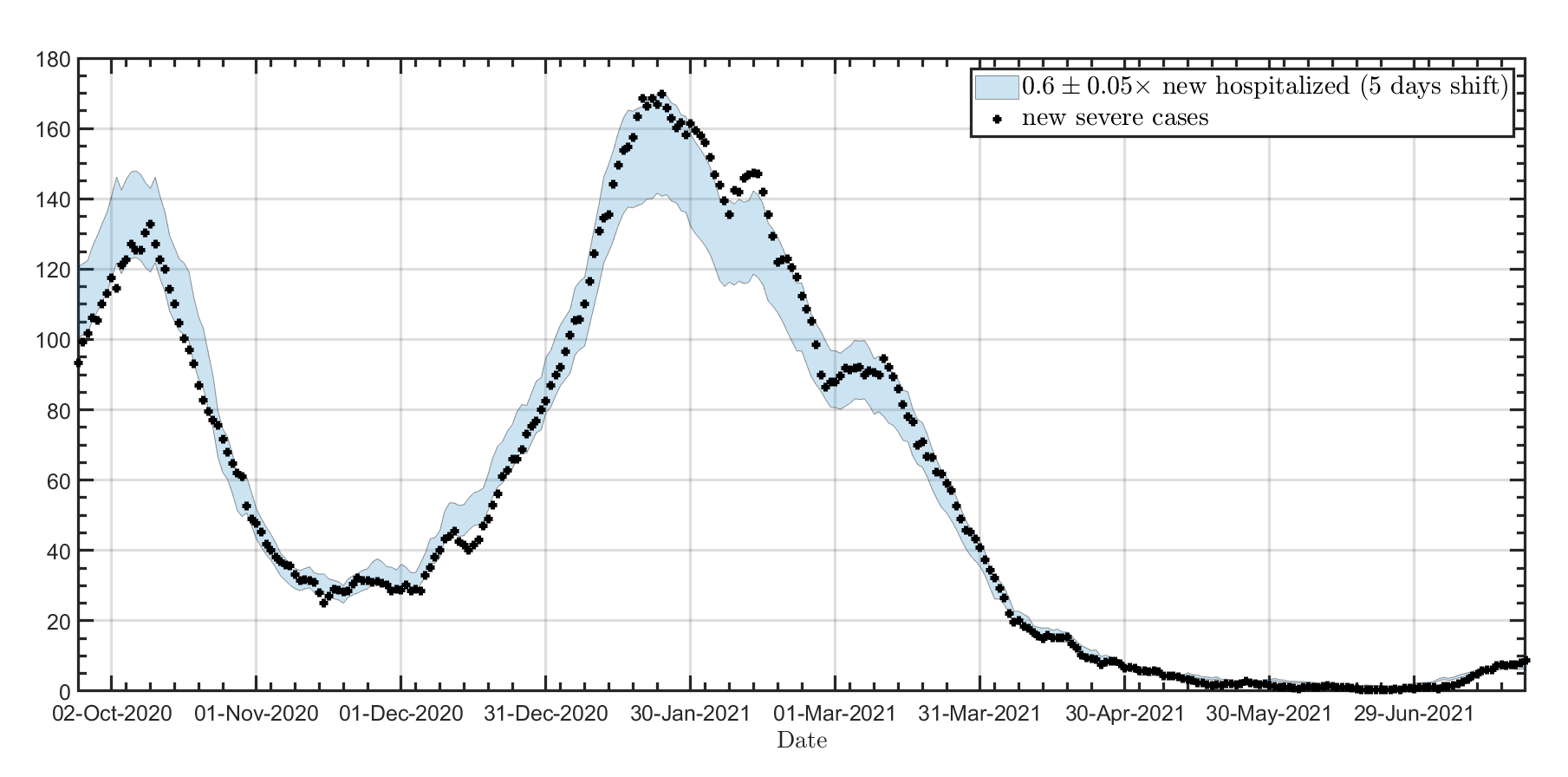}
\caption{\footnotesize{The correlation between the new daily severe and moderate hospitalized and the new severe hospitalized after five days. Dots: new severe hospitalized; Solid blue band: $0.6\pm 0.05 \times$ the new severe and moderate hospitalized with five days shift (all the data was taken from Refs.~\cite{dancarmoz,datagov}}}
\label{fig_severe_SM}
 \end{figure}

 Also, based on the current data from Israel, we found that there is a 5-days shift from being hospitalized to becoming a severe case (Fig.~\ref{fig_severe_SM}), where the number of the new daily severe cases is approximately 0.6 of the new hospitalized five days earlier (all the data was taken from Refs.~\cite{dancarmoz,datagov}).
\subsubsection{Simulating the time dependence of severe cases and age distribution}
\begin{figure}[h!]
 \centering
 \includegraphics[width=1\linewidth]{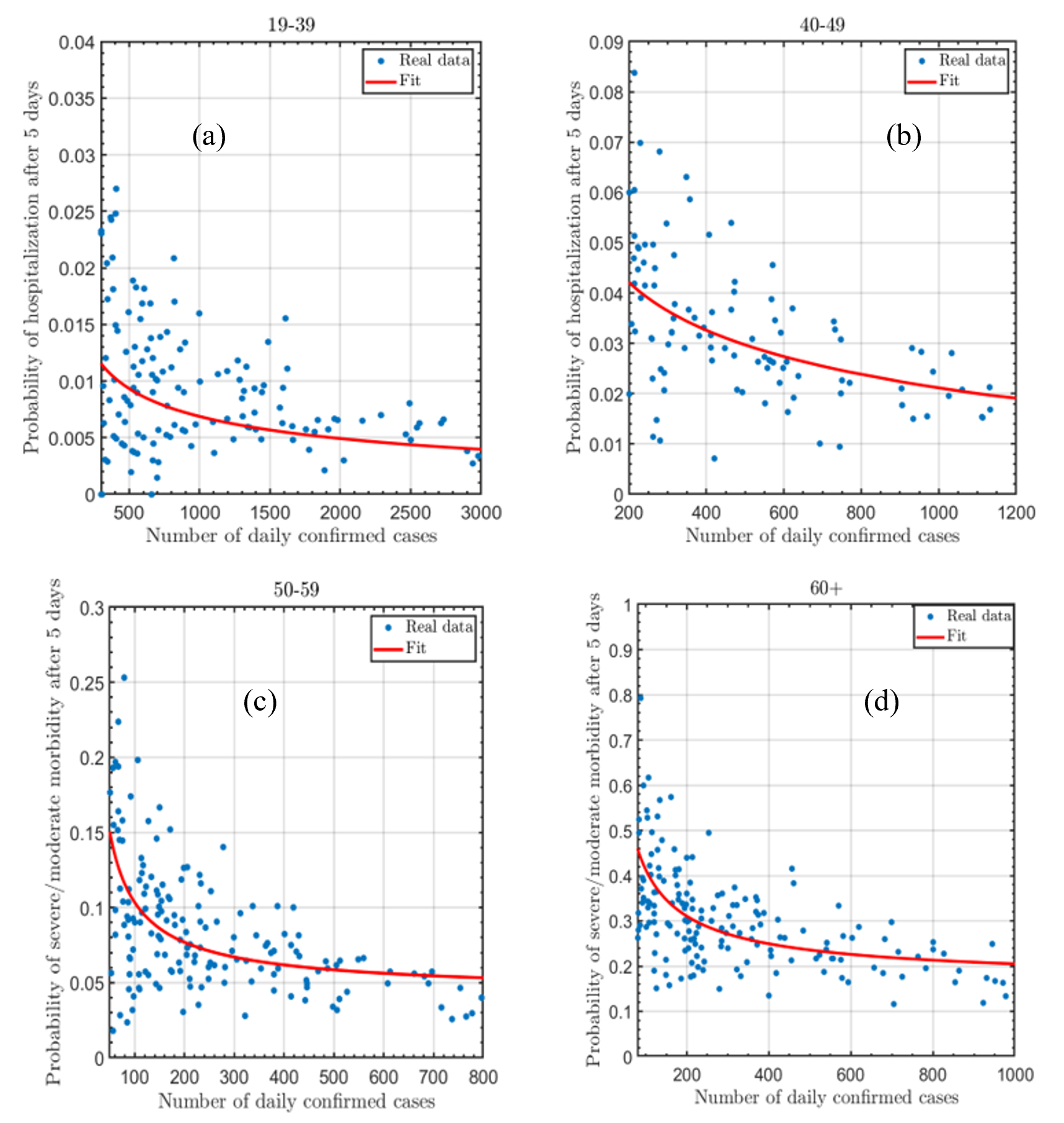}
 \caption{\footnotesize{the probability of moderate and severe morbidity as a function of the daily confirmed cases (with a five-day shift) for four age groups; (a):19-39;(b) 40-49;(c) 50-59 (d) 60+. For all panels, the dots represent real data of the moderate and severe morbidity in Israel between 01/08/2020 and 31/10/2020. The solid line is fitted function}}
 \label{fig:SM_ages}
\end{figure}

The inputs $R_e$ and $R_t$ allow a prediction of the number of confirmed cases. As shown by using Fig.~\ref{fig:SM_cases}, this leads to admissions in moderate or severe condition (and as a result, the number of new severe cases as shown in Fig.~\ref{fig_severe_SM}). We then calibrated the simulation to reproduce the new moderate and severe cases around the second lockdown period (i.e., between August 1, 2020, and October 31, 2020). The calibration process leads to a fit of the probability of a person being hospitalized in this condition. It is found that the probability of hospitalization decreases as the load on hospitals increases. This is a measure of the effect of hospital capacity on the treatment an individual receives. 
Fig.~\ref{fig:SM_ages} presents the percentage of moderate and severe morbidity as a function of the daily confirmed cases (with a five-day shift) for four age groups from 19 and up (where the groups were distinguished as mentioned in Section A). Data and their fitted analytical functions are presented for elderly (over 60) and younger (under 60) patients, the severe/moderate morbidity in the current wave can be estimated for both age groups. These functions are then used in the paper to predict severe  morbidity using Figure.~\ref{fig_severe_SM}. 

\bibliography{ref}
\bibliographystyle{unsrt}